\documentclass[twoside,12pt]{article}
\usepackage{epsfig}

\newcommand{\be}{\begin{equation}}
\newcommand{\ee}{\end{equation}}
\newcommand{\bea}{\begin{eqnarray}}
\newcommand{\eea}{\end{eqnarray}}

\begin{document}
\title{ \vspace{1cm}Low-Energy Nuclear Astrophysics  - the Fascinating Region of
A=7}
\author{Michael Hass\\
Department of Particle Physics, The
Weizmann Institute of Science Rehovot, Israel}
\maketitle
\begin{abstract}
We discuss results and future plans for low-energy reactions that
play an important role in current nuclear astrophysics research and
that happen to concentrate around the region of A=7. The
$^7$Be(p,$\gamma$)$^8$B and the $^3$He($^4$He,$\gamma$)$^7$Be
reactions are crucial for understanding the solar-neutrino
oscillations phenomenon and the latter one plays a central role in
the issue of cosmic $^7Li$ abundance and Big-Bang Nucleosynthesis.
We also present results regarding the host dependence of the half
life of the electron-capture  $^7$Be radio-nuclide.
\end{abstract}
\section{SOLAR FUSION REACTIONS, SOLAR-NEUTRINOS and BIG-BANG NUCLEOSYNTHESIS}

\subsection{The cross section of the $^7$Be(p,$\gamma$)$^8$B reaction}
    Our planet is bombarded every second with a large number of charge-less, mass-less neutrinos,
    originating in the nuclear fusion reactions that power the energy production in the Sun. A major,
    long-lasting research effort, focusing on the apparent shortfall of detected solar neutrinos as
    compared to theoretical predictions, culminated recently with the results of the Sudbury Neutrino
    Observatory (SNO) experiment \cite{SNO}, affirming the notion of neutrino oscillations
    (and hence neutrino mass). The SNO experiment, as well as the preceding Homestake \cite{Cleveland}
    and Super-Kamiokande \cite{Fukuda} experiments are sensitive only to a small fraction of the solar
    neutrino spectrum, to the high-energy neutrinos that are emitted as a result of the fusion of
    protons with the nucleus $^7$Be and the subsequent beta decay of the $^8$B product nucleus.
    The cross section of this reaction has been measured in the laboratory several times.
    However, mostly due to difficulties with the preparation and homogeneity of the
    radioactive $^7$Be target, large discrepancies still persist in the extracted cross
    section values. The present experiment uses in a novel way a 2 mm diameter target
    of the $^7$Be radioactive nuclei (with a halflife of 53 days), prepared by direct
    implantation at the ISOLDE (CERN) laboratory and brought to the Van de Graaff
     accelerator of the Weizmann Institute, Israel, for the measurement of the reaction.
     The results of these experiments have been published in detail in several previous
     papers \cite{Weissman,Hass,Baby,Baby1,Baby2}. Fig. 1 presents a brief summary of the results.

\begin{figure}[ht]
\vskip.2in \centering
\includegraphics[width=0.6\textwidth]{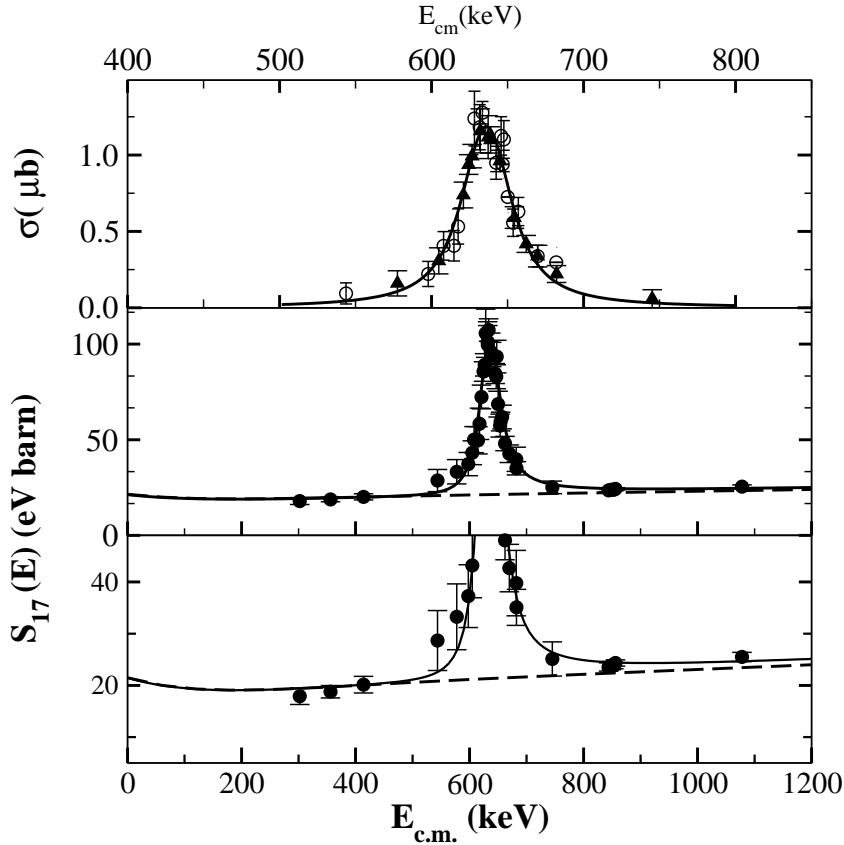}
\caption{The astrophysical S$_{17}$ factor (the cross section of the
reaction, with the penetrability through the Coulomb barrier
factored out) from the present measurement. The extrapolation to E=0
yields S$_{17}$(0)=21.5 (7) eV.b. Depicted in the figure are the
regions of the M1 resonance and those above and below it.}
\end{figure}

With the present determination of the p+$^7$Be cross section at
solar energies to an
 accuracy of better than 4\%, this important nuclear physics quantity ceases to be the
 largest source of error in the Standard Solar Model \cite{Bahcall} estimates of the $^8$B
 neutrino flux.

\subsection{The cross section of the $^3$He($^4$He,$\gamma$ )$^7$Be reaction}
    The $^3$He($^4$He,$\gamma$ )$^7$Be reaction is another major source of uncertainty in determining the solar
    neutrino flux that is proportional approximately \cite{Bahcall} to the astrophysical S factor of this
    reaction, S$_{34}$(0)$^{0.8}$. A compilation of previous data, together with the latest measurement by the authors, is presented by Hilgemeier {\it et al.} in Ref.\cite{Hilgemeier}, from which a significant scatter in  S$_{34}$(0) can be seen, resulting in a considerable uncertainty on the adopted S$_{34}$(0). It is therefore highly important to provide a more precise value than the presently recommended value of S$_{34}$(0) = 0.51$\pm$0.02 keV-b \cite{Hilgemeier}. We have recently initiated a new precision measurement of this cross section at energies around E$_{c.m.}$ = 500 - 950 keV to serve as a normalization to the S$_{34}$(E) curve, using a $^3$He beam and a $^4$He gas cell and determining the cross section by measuring the ensuing $^7$Be activity. Since the energy dependence among various theoretical approaches and experimental determinations agrees quite well \cite{Tombrello, Williams}, the normalized curves will then be compared with S(E) values from our ongoing experiment to obtain S$_{34}$(0).
We note here also our earlier attempt towards a new determination of
S$_{34}$(0) by using the accelerator mass spectrometry technique
\cite{Bordeanu}.

A schematic diagram of our experimental setup is shown in Fig. 2.
The $^3$He beam from the 3 MV Van de Graaff accelerator at the
Weizmann Institute of Science enters the $^4$He gas cell through a
nickel (Ni) window of either 0.5 or 1 $\mu$m and is raster-scanned
in order to avoid excessive localized heating of the Ni window.  The
beam direction is defined by an upstream slit at 2 meters from the
center of the cell and two Ta collimators of 3 mm, one at the
entrance and the other at the exit of the chamber. An aperture
operated at -400 V, placed before the Ni window, serves as a
secondary electron suppressor. The gas cell is insulated from the
beam line and the entire chamber, including the Cu stopper that is
in electric contact with the chamber, servs as Faraday cup to
determine the number of impinging $^3$He particles.  $^7$Be nuclei
produced by fusion of $^3$He on $^4$He in the cell move forward in
the laboratory system and are implanted in the Cu catcher at depths
of few microns.   The catcher is kept at distances of 10.3 or 13.9
cm from the Ni foil and the $^4$He gas pressure inside the cell is
accordingly adjusted to obtain $\sim$120µg/cm$^2$ thickness of gas.
This corresponds to an energy width of the target of $\sim$200 keV
at E$_{cm}$ = 1.1 MeV. The gas pressure was monitored and maintained
at a constant pressure. In addition, we monitor on-line the
elastically scattered $^3$He from the Ni window, using a Si surface
barrier detector placed at 44.68$^{\circ}$ with respect to the beam
direction. This provides a crosscheck of the beam current
measurement.

\begin{table}[htb]
\caption{Preliminary results from the present experiment at
E$_{c.m.}$ = 952 keV. The cross section and S-factor values are
quoted in arbitrary units for both current integration (CI) and
elastic scattering (Rut.) beam-particle normalizations as described
in the text. The total number N($^7$Be, t=0) of $^7$Be nuclei is
normalized to correspond to the end of the implantation.}
\label{table:1}
\newcommand{\m}{\hphantom{$-$}}
\newcommand{\cc}[1]{\multicolumn{1}{c}{#1}}
\renewcommand{\tabcolsep}{.5pc} 
\renewcommand{\arraystretch}{1.2} 
\begin{tabular}{@{}llllll}
\hline

E$_{c.m.}$  &N($^7$Be, t=0)&$\sigma_{CI}$&    $\sigma_{Rut}$& S$_{}$ (E)&S$_{Rut}$(E)\\
&&[arb. units]& [arb. units]& [arb. units] &
[arb. units]\\
\hline
952 &6.64 10$^6$ (3.5\%) &1000    &1000    &0.400   &0.400\\
952 &8.61 10$^6$ (3.0\%) &911     &893     &0.364   &0.357\\
952 &6.89 10$^6$ (4.1\%) &1024    &1089    &0.409   &0.434\\
952 &5.86 10$^6$ (4.5\%) &952     &1012    &0.380   &0.405\\
951 &10.4 10$^6$ (3.6\%) &1006    &940     &0.402   &0.377\\

\hline

\end{tabular}\\[2pt]
\end{table}

\begin{figure}[ht]
\vskip.2in \centering
\includegraphics[width=0.6\textwidth]{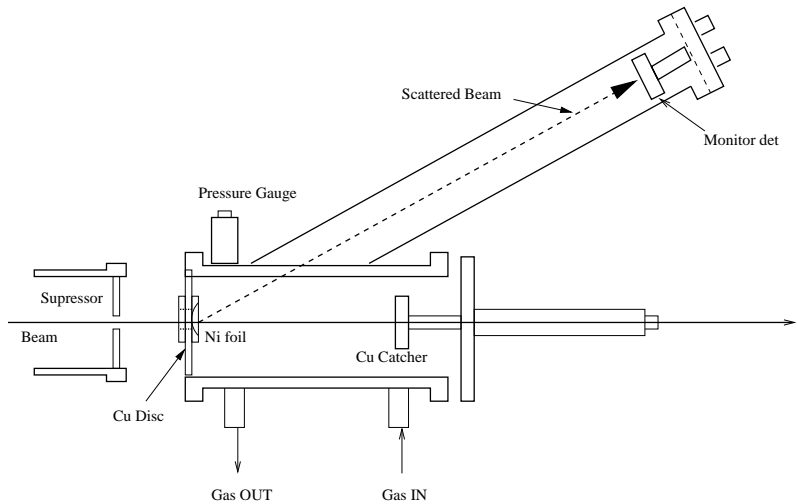}
\caption{A schematic drawing of the chamber. The various items
described above, such as the gas foil, electron suppression, the Si
detector and the Cu stopper, are depicted.}
\end{figure}
The $^7$Be nucleus decays by a 478 keV  $\gamma$-ray transition from
the first excited state in $^7$Li with a lifetime of T$_{1/2}$=
53.29 $\pm$ 0.07 d and a branching ratio of 10.45 $\pm$ 0.04\% [see,
e.g., Refs. 7,8, and references therein]. The number of $^7$Be
nuclei produced in the reaction is obtained by measuring the 478 keV
$\gamma$-activity at the NRC-Soreq laboratory, using a large-volume
Ge detector, placed in a specially shielded environment for low
count rate measurements. A similar setup was used previously by our
group to determine with high accuracy the number of $^7$Be atoms in
our implanted $^7$Be target for the determination of S$_{17}$(0)
\cite{Baby, Baby1}. A typical $\gamma$-ray spectrum is shown in Fig.
3. The results from our recent experiment \cite{Nara} are shown in
Table I and in Fig. 4.

\begin{figure}[ht]
\vskip.2in \centering
\includegraphics[width=0.6\textwidth]{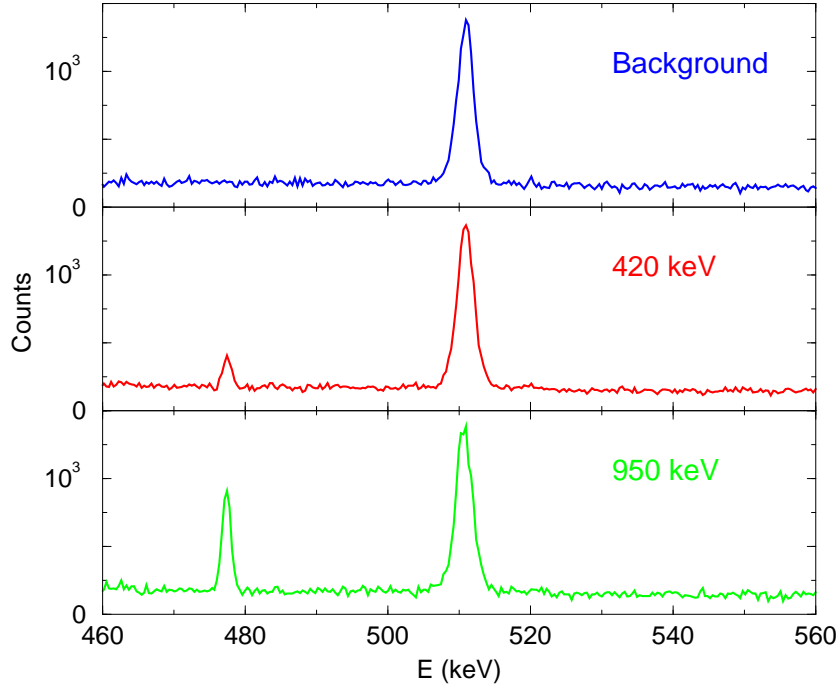}
\caption{A typical spectrum from the decay of  6$\cdot$106 atoms of
$^7$Be, corresponding to one of the measurements presented in Table
I. The 478 keV $\alpha$-peak can be seen prominently with very small
background.}\label{Fig.3}
\end{figure}

\begin{figure}[!ht]
\vskip.2in \centering
\includegraphics[width=10cm]{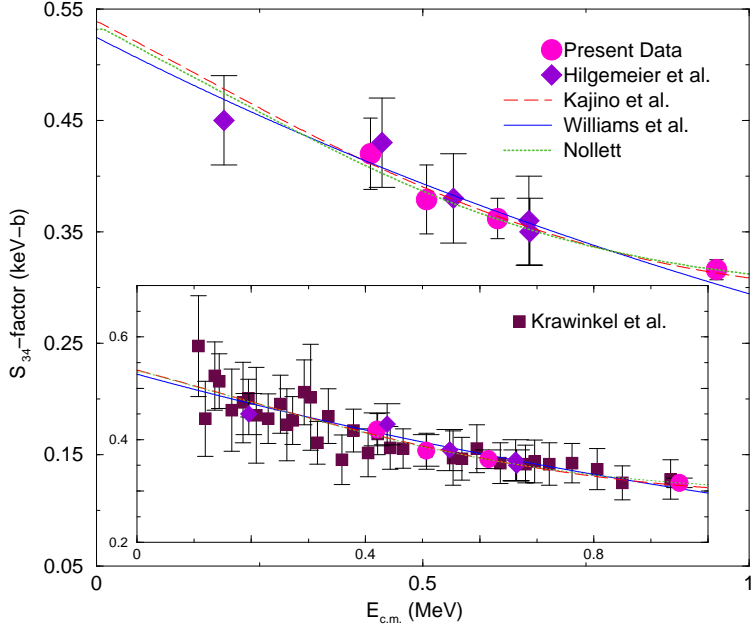}
\hfill \caption{Present data together with that of  previous results
and representative theoretical fits, yielding $S_{34}$(0)~=~0.533
(20)(7)~keV-b.} \label{Fig.4}
\end{figure}

The value of the S factor we obtain is S$_{34}$(0) = 0.53(2)(1)
keV.b, in excellent agreement with earlier compilation but with a
much reduced error. Our present result sis also in excellent
agreement with a recent experiments of the Luna collaboration
(\cite{Luna}).

\section{HOST-DEPENDENCE of the $^7$Be HALF LIFE}
The decay rate of radioactive nuclei that undergo orbital Electron
Capture (EC) depends on the properties of the atomic electron cloud
around the nucleus. Hence, EC may exhibit varying decay rates if the
nucleus is implanted into  host materials with different properties
of their corresponding electron clouds. The first suggestion of this
effect in $^{7}$Be, which is the lightest nucleus that decays by EC,
and reports of experiments trying to investigate this phenomenon,
have been presented by Segr\`{e} et al.~\cite{Segre,SegreWie,Lein}.
This effect has been qualitatively attributed in the past to the
influence of the electron affinities of neighboring host atoms
~\cite{Das}. The electron density of the $^{7}$Be atom in a
high-electron affinity material such as gold is decreased via the
interaction of its 2s electrons with the host atoms, resulting in a
lower decay rate (longer half-life). Recently, the life-time
modification has been suggested to stem from differences of the
Coulomb screening potential ~\cite{Kettner} between conductors and
insulators (see below).

Several experimental and theoretical investigations were conducted
during recent years to study the host material effect on the decay
rate of $^{7}$Be ~\cite{Das,Ray,Norman,Ray1,Ohtsuki,Ray2}, with a
somewhat confusing scattering of experimental results. It was found
that the half-life of $^{7}$Be encapsulated in a fullerene C$_{60}$
cage and $^{7}$Be in Be metal is 52.68(5) and 53.12(5) days,
respectively, amounting to a difference of 0.83(13)\%
\cite{Ohtsuki}. A smaller effect of $\approx$0.2\% was measured for
the half-life 53.64(22) d of $^{7}$Be in C$_{60}$ and 53.60(19) d of
$^{7}$Be in Au ~\cite{Ray2}. A recent theoretical evaluation shows
that short and long half-lives 52.927(56)d and 53.520(50)d were
measured for $^{7}$Be in Al$_{2}$O$_{3}$ and $^{7}$Be in (average of
BeO, BeF$_{2}$ and Be(C$_{5}$H$_{5}$)$_2$), respectively
~\cite{Das}. These results yield the magnitude of the effect to be
as large as 1.1$\%$. Another experimental investigation has shown
that the half-life increases by 0.38$\%$ from $^{7}$Be in graphite,
53.107(22) d, to $^{7}$Be in Au, 53.311(42) ~\cite{Norman}, while a
very recent investigation ~\cite{Limata} has seen no effect to
within 0.4\%. The great interest in this phenomenon for $^7$Be
arises also from the need to explore the possible contribution of
the half-life of $^{7}$Be to the measurement of the cross section of
the two fusion reactions, $^{7}$Be(p,$\gamma$)$^{8}$B and
$^{3}$He($^{4}$He,$\gamma$)$^{7}$Be, that play an important role in
determining the solar neutrino flux ~\cite{Baby,Nara}.

The present work has been undertaken in order to probe this
phenomenon yet again in an experimental approach that takes full
advantage of the experience gained in measuring implanted $^{7}$Be
activity in a controlled and precise manner for cross section
determinations of the solar fusion reactions mentioned above
~\cite{Baby,Nara}. As a demonstration of the quality of the
$\gamma$-activity measurement, we cite the results of ~\cite{Baby,
Baby1} where two independent determinations of the $\it{absolute}$
activity of $^7$Be, at the Soreq laboratory and at Texas A$\&M$
University, were in excellent agreement to within 0.7$\%$.  The same
setup has also been used for determining the $^7$Be activity ensuing
from the $^{3}$He($^{4}$He,$\gamma$)$^{7}$Be reaction ~\cite{Nara}.
We report the measurement of the half-life of $^{7}$Be implanted in
four host materials: copper, aluminum, aluminum oxide (sapphire -
Al$_{2}$O$_{3}$) and PVC (polyvinyl chloride - [C$_2$H$_3$Cl]$_n$]).

The primary source of $^{7}$Be for implantation was a graphite
target, from the Paul Scherrer Institute (PSI), used routinely for
the production of $\pi$ mesons ~\cite{Heid}. Many spallation
products are accumulated in the target, including $^7$Be. Graphite
material from the PSI meson production target was placed in an
ion-source canister and was brought to ISOLDE (CERN); $^7$Be was
extracted at ISOLDE by selective ionization using a resonance laser
ion source. Direct implantation of $^{7}$Be at 60 keV in the host
material was subsequently followed. A detailed description of the
extraction and implantation of $^7$Be at ISOLDE is provided in
detail in Refs. ~\cite{Baby1,Kost}. This procedure facilitated a
precision measurement of the cross section of the reaction
$^{7}$Be(p,$\gamma$)$^{8}$B. The implantation spot was defined by a
2 mm collimator positioned at close proximity to the target for the
Cu sample and a 5 mm collimator for the other samples. This small
change in the ensuing counting geometry has been well investigated
for the measurement of $^7$Be activity ~\cite{Nara} and does not
affect the results in any significant manner. The implantation
process provided full control of the spot composition ($^{7}$Be;
$^{7}$Li) as well as a radial and depth profiles. For earlier
implantations, at a density of $^7$Be in Cu far exceeding that of
the present experiment, the spot was found to be robust and the
$^{7}$Be inventory in the spot was stable ~\cite{Baby1,Hass},
excluding naturally radioactive decay. The copper, aluminum and PVC
host material targets consisted of disks of 12 mm diameter and 1.5
mm thickness, while the sapphire target was a square of 10.2 mm x
10.2 mm. The median implantation depth of $^{7}$Be into these
materials has been estimated using the SRIM code ~\cite{Srim} and
found to be 12, 24, 470 and 37 $\mu$m, respectively, i.e. all
implantation depths were well below the surface.

A typical decay curve is presented in Fig. 5  and the results of the
four host materials, Al and Cu (conductors) and  PVC and Al$_2$O$_3$
are presented in Fig. 6. Even though the statistical test of the
present data supports a null effect within $\pm$ 1$\sigma$, the
results of Fig. 5 may indicate a slight positive trend of the
half-life versus the electron affinity, where a host material with
high-electron affinity such as copper (conductor), exhibits a longer
half-life, compared to a lower electron affinity material such as
aluminum oxide (insulator).

A possible interpretation of the life-time results follows a recent
observation by Wang et al. ~\cite{Wang} of an approximately 1$\%$
increase in the lifetime of $^7$Be in metallic vs. insulator
environments at low temperature. This change is consistent with the
Debye screening model ~\cite{Kettner} that has been successfully
used to explain the screening potential for nuclear reactions at
very low beam energies. The average life times for the two
insulators (PVC and Al$_2$O$_3$) and the two metals (Cu and Al) are
53.180(31) d and 53.299(33) d, respectively, a difference of
0.22$\%$. Indeed, the trend of the present data is in basic
agreement with the temperature dependence of the screening model, as
well as with the results of ~\cite{Limata}. A further investigation
of such a small trend and its detailed temperature dependence is
clearly called for.

\begin{figure}[!h]
\vskip.2in \centering
\includegraphics[width=0.7\textwidth]{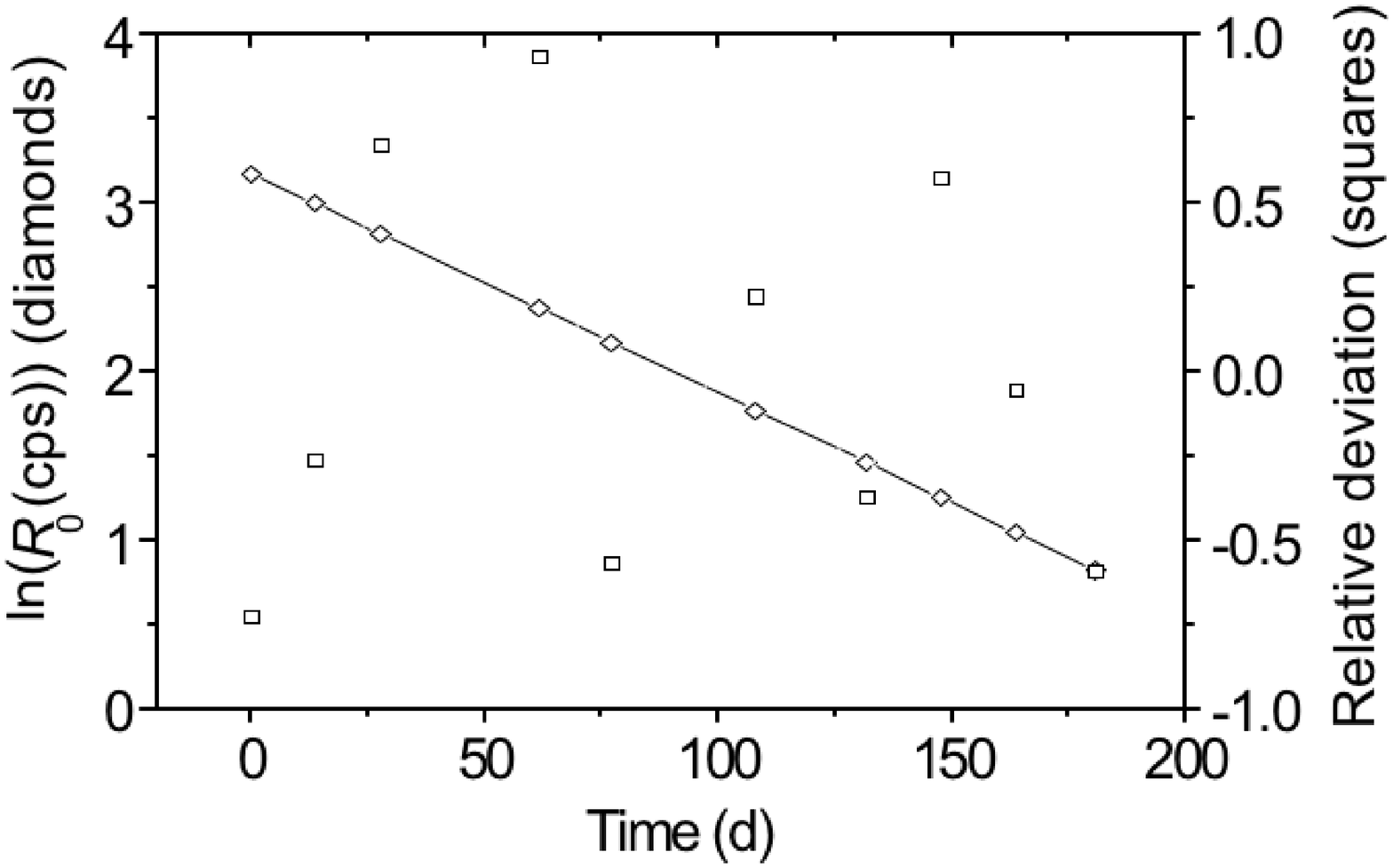}
\hfill \caption{Top (left axis, diamonds): The decay curve of the
477.6 keV line of $^7$Be imbedded in the Al$_2$O$_3$ sample. The
straight line was fitted by a weighted linear regression. Bottom
(right axis, squares): The deviation between the measured and the
fitted count rates, divided by the corresponding uncertainty, for
the 10 measurements of the Al$_2$O$_3$ sample (see text).}
\label{Fig.5}
\end{figure}

\begin{figure}[!h]
\vskip.2in \centering
\includegraphics[width=0.6\textwidth]{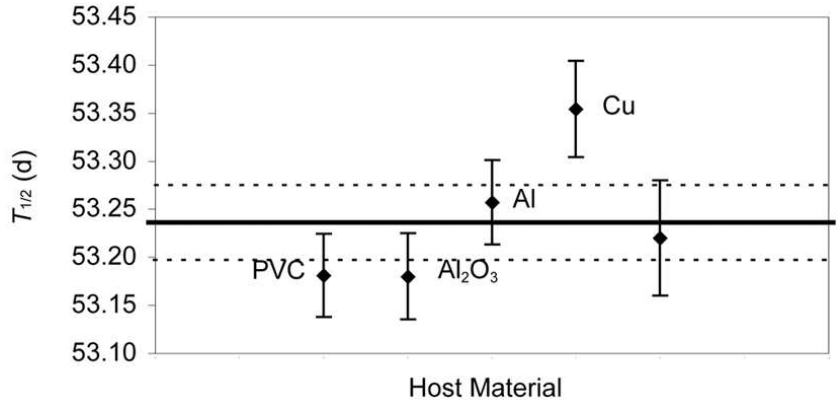}
\hfill \caption{The half life of $^7$Be in 4 host materials. The
solid line represents the weighted average and the broken lines
correspond to a $\pm$1$\sigma$ interval. Also shown is the adopted
value in the literature ~\cite{Trilley}} \label{Fig.6}
\end{figure}

\section{Acknowledgments:}
I would like to thank my close co-workers at the Weizmann Institute
and all my other colleagues for a fruitful collaboration. We
acknowledge the support of the Israel Science Foundation (ISF).


\begin{thebibliography}{9}
\bibitem{SNO}SNO collaboration, Phys. Rev. Lett. {\bf87}, 0713101 (2001);
Phys. Rev. Lett. {\bf89}, 011301 (2002).
\bibitem{Cleveland}B.T. Cleveland et al., J. Exp. Theor. Phys. {\bf496}, 505 (1998).
\bibitem{Fukuda}S. Fukuda et al., Phys. Rev. Lett. {\bf86}, 5651 (2001);
\bibitem{Weissman}L. Weissman, C. Broude, G. Goldring, R. Hadar, M. Hass,
F. Schwamm and M. Shaanan, Nucl. Phys. {\bf A630}, 678 (1998).
\bibitem{Hass}M. Hass, C. Broude, V. Fedoseev, G. Goldring, G. Huber,
J. Lettry, V. Mishin, H.J. Ravn, V. Sebastian and L. Weissman,
Phys. Lett. {\bf B462}, 237 (1999).
\bibitem{Baby}L.T. Baby, C. Bordeanu, G. Goldring, M. Hass, V. Fedoseev,
U. Koester, L. Weissman, Y. Nir-El, G. Haquin,  H. W. Gaggeler, R. Weinreich
and the ISOLDE collaboration Phys. Rev Lett. {\bf 90}, 22501 (2003).
\bibitem{Baby1}L.T. Baby, C. Bordeanu, G. Goldring, M. Hass, V. Fedoseev,
U. Koester, L. Weissman, Y. Nir-El, G. Haquin, H. W. Gaggeler, R. Weinreich
and the ISOLDE collaboration, Phys. Rev. {\bf C67}, 065805 (2003).
\bibitem{Baby2}L.T. Baby, C. Bordeanu, G. Goldring, M. Hass, V. Fedoseev,
U. Koester, L. Weissman, Y. Nir-El, G. Haquin, H. W. Gaggeler, R. Weinreich
and the ISOLDE collaboration,  Nucl. Phys. {\bf A718}, 477c (2003).
\bibitem{Bahcall}See, e.g., J.N. Bahcall, M. Pinsonneault and S. Basu,
Astophys. J.  {\bf555}, 990 (2001), and references therein.
\bibitem{Hilgemeier}M. Hilgemeier et al., Z. Phys. {\bf A329}, 243 (1988).
\bibitem{Tombrello}T.A. Tombrello et al., Phys. Rev. {\bf 131}, 2582 (1963).
\bibitem{Williams}R.D. Williams et al., Phys. Rev. {\bf C23}, 2773 (1981).
\bibitem{Nara} B.S. Nara Singh et al., Phys. Rev. Lett. {\bf 93, 262503
(2004)}
\bibitem{Luna} D. Bemmerer et al., Phys. Rev. Lett. {\bf 97, 122502
(2006)}.
\bibitem {Segre}E. Segr\`{e}, Phys. Rev. {\bf 71}, 274 (1947).
\bibitem {SegreWie}E. Segr\`{e} and C. E. Wiegand, Phys. Rev. {\bf 75}, 39 (1949).
\bibitem {Lein}R.F. Leininger et al., Phys. Rev. {\bf 76}, 897 (1949).
\bibitem {Das}P. Das and A. Ray, Phys. Rev. C{\bf 71}, 025801 (2005).
\bibitem {Kettner} K.U. Kettner et al., J. Phys. G {\bf 32}, 489 (2006);
and references therein.
\bibitem {Ray}A. Ray et al., Phys. Lett. B{\bf 455}, 69 (1999).
\bibitem {Norman}E.B. Norman et al., Phys. Lett. B{\bf 519}, 15 (2001).
\bibitem {Ray1}A. Ray {\it et al.}, Phys. Lett. B{\bf 531}, 187 (2002).
\bibitem {Ohtsuki}T. Ohtsuki et al., Phys. Rev. Lett. {\bf 93}, 112501 (2004).
\bibitem {Ray2}A. Ray et al., Phys. Rev. C{\bf 73}, 034323 (2006).
\bibitem {Limata} B.N. Limata et al., Eur. Phys. J. A{\bf 27}, s01,
193-196 (2006).
\bibitem {Heid} G. Heidenreich et al., Proc. AIP {\bf 642}, 122
(2002 ).
\bibitem {Kost} U.~K\"{o}ster et al., Nuc. Inst. Meth. B{\bf
204}, 343 (2003).
\bibitem {Hass} M. Hass et al., Phys. Lett. B{\bf 462}, 237 (1999).
\bibitem {Srim}SRIM package from www.srim.org
\bibitem {Wang} B. Wang et al., Eur. Phys. J. A{\bf 28}, 375-377;
and references therein (2006).
\bibitem {Kettner} K.U. Kettner et al., J. Phys. G {\bf 32}, 489 (2006);
and references therein.
\bibitem {Limata} B.N. Limata et al., Eur. Phys. J. A{\bf 27}, s01,
193-196 (2006).

\end{thebibliography}
\end{document}